\documentclass[preprint]{aastex}
\usepackage{emulateapj5}
 
\newcommand{\kms}{{\ifmmode{{\rm km~s}^{-1}}\else{km~s$^{-1}$}\fi}}
\newcommand{\delv}{{\ifmmode{\Delta v}\else{$\Delta v$}\fi}}
\newcommand{\cm}{{\ifmmode{{\rm cm}^{-1}}\else{cm$^{-1}$}\fi}}
\newcommand{\cmm}{{\ifmmode{{\rm cm}^{-2}}\else{cm$^{-2}$}\fi}}
\newcommand{\cmmm}{{\ifmmode{{\rm cm}^{-3}}\else{cm$^{-3}$}\fi}}

\newcounter{species} 
\def\ion#1#2{\setcounter{species}{#2}#1$\;${\scriptsize\Roman{species}}\relax}

\slugcomment{\today}
\shorttitle{New NIR DIBs in the Orion Nebula}
\shortauthors{Misawa et al.}

\begin{document}

\title{Identification of New Near-Infrared DIBs in the Orion
Nebula\altaffilmark{1}}

\footnotetext[1]{Based on data collected at Subaru Telescope, which is
operated by the National Astronomical Observatory of Japan.}

\author{Toru Misawa\altaffilmark{2},
        Poshak Gandhi\altaffilmark{2},
        Akira Hida\altaffilmark{3},
        Toru Tamagawa\altaffilmark{2}, and
        Tomohiro Yamaguchi\altaffilmark{3}}

\altaffiltext{2}{Cosmic Radiation Laboratory, RIKEN, 2-1 Hirosawa,
  Wako, Saitama 351-0198 Japan}
\altaffiltext{3}{Advanced Device Laboratory, RIKEN, 2-1 Hirosawa,
  Wako, Saitama 351-0198 Japan}

\email{misawa@crab.riken.jp}

\begin{abstract}
Large organic molecules and carbon clusters are basic building blocks
of life, but their existence in the universe has not been confirmed
beyond doubt.  A number of unidentified absorption features (arising
in the diffuse inter-stellar medium), usually called ``Diffuse
Inter-stellar Bands (DIBs)'', are hypothesized to be produced by large
molecules. Among these, buckminsterfullerene $C_{60}$ has gained much
attention as a candidate for DIB absorbers because of its high
stability in space.  Two DIBs at $\lambda$ $\sim$ 9577~\AA\ and
9632~\AA\ have been reported as possible features of $C_{60}^+$.
However, it is still not clear how their existence depends on their
environment.  We obtained high-resolution spectra of three stars
in/around the Orion Nebula, to search for any correlations of the DIB
strength with carrier's physical conditions, such as dust-abundance
and UV radiation field.  We find three DIBs at $\lambda$ $\sim$
9017~\AA, 9210~\AA, and 9258~\AA\ as additional $C_{60}^+$ feature
candidates, which could support this identification.  These DIBs have
asymmetric profiles similar to the longer wavelength features.
However, we also find that the relative strengths of DIBs are close to
unity and differ from laboratory measurements, a similar trend as
noticed for the 9577/9632 DIBs.
\end{abstract}


\section{Introduction}
The Diffuse Interstellar Bands (hereafter, DIBs) are a series ($\sim$
300) of broad absorption lines at $\lambda$ = 4000\AA\ to 1.3$\mu$m.
Since the first detections \citep{heg22,mer34}, various molecules,
including organic compounds (e.g., polycyclic organic hydrocarbons,
fullerenes, carbon chains, and carbon nanotubes), have been proposed
as the source of DIBs, because they are all relatively stable within
the harsh radiation environment of space.  Nonetheless, it has not
been confirmed observationally what the {\it real} source(s) of DIBs
are/is, and this remains one of the most long-standing unsolved
questions in Astronomy.

Originally, DIBs were detected in the inter-stellar medium (ISM)
toward Galactic stars (e.g., \citealt[]{her95}), and then were also
detected in the Large and Small Magellanic Clouds (e.g.,
\citealt[]{cox05}), in nearby galaxies \citep{sol05}, in starburst
galaxies \citep{hec00}, and in high column density \ion{H}{1}
absorption systems (i.e., Damped Ly$\alpha$ systems) at redshifts up
to $z$ $\sim$ 0.5 \citep{yor06}. Thus, DIB absorbers are ubiquitous in
space.

Among $\sim$300 DIBs, only a few have specific source candidates:
e.g., large ionized $C_{60}$ fullerene ($C_{60}^{+}$) at $\lambda$ =
9577~\AA\ and 9632~\AA\ \citep{foi94}, several fullerenes ($C_{80}$,
$C_{240}$, $C_{320}$, $C_{540}$) at $\lambda$ = 4430~\AA\
\citep{igl07}, and naphthalene cation ($C_{10}H_{8}^{+}$) at $\lambda$
= 6125~\AA, 6489~\AA, and 6707~\AA\ \citep{igl08}.  These DIBs have
wavelengths that are very close to those measured in laboratory.  By
running a variety of gas-phase chemical models, \citet{bet96}
confirmed that linear hydrocarbons are spontaneously converted into
cyclic rings and fullerenes.  Among these, the DIBs at $\lambda$ =
9577~\AA\ and 9632~\AA\ are most clearly detected, and have been
discussed frequently (e.g., \citealt{foi94,foi97,gal00}).  However,
their identification is still in question, because \citet{foi94} did
not report on the other weaker transitions of $C_{60}^{+}$ at
$\lambda$ = 9366~\AA\ and 9419~\AA, which should exist in conjunction
with the two detected DIBs, if their carrier really is $C_{60}^{+}$
\citep{jen97}.  Further observations that would support this
identification are required. In this paper, we concentrate on the ISM
in the Orion Nebula, and report on the detectability of several DIBs
at 9000~\AA\ -- 9700~\AA, depending on the stellar environments.

\section{Observations and Data Reduction}
We obtained spectra of three O/B stars in or around the Orion Nebula
on 2008 November 14 (UT) with the Subaru telescope and High Dispersion
Spectrograph (HDS; \citealt[]{nog02}).  The positions of our targets
are shown in Figure~1.  Two stars (HD~37022 and HD~37041) lie in the
central region of the nebula.  The other star (HD~37150) is located at
the edge of the nebula.  Because our targeted wavelength region is at
$\sim$ 9000~\AA\ -- 10000~\AA, a number of telluric lines give severe
contamination.  Therefore, we observed a standard star
(HD~37397)\footnote{This is a rapid rotator, located close to the
Orion Nebula.  Its high rotational velocity ($v\sin i$ $>$ 350~\kms)
and spectral type earlier than B2 \citep{gle00}, are optimal for our
purpose.} near the nebula to divide out these lines, as described in
Section 2.1.  Stellar intrinsic features are smoothed out, because
HD~37397 is a rapid rotating star with rotational velocity of $v\sin
i$ = 355~\kms\ \citep{gle00}.  We used a 0{\farcs}3 slit width ($R =
120,000$), and adopted 2 pixel binning only along the spatial
direction.  The red grating, with a central wavelength of 8650~\AA,
was used to cover the wavelength of 7310~\AA\ and 8560~\AA\ on the
blue CCD chip and 8710\AA\ and 10000\AA\ on the red CCD chip.  Only
the red CCD data was used because the blue CCD photon counts were
saturated and deviated from linearity.  We reduced the data in a
standard manner using the IRAF software\footnote{IRAF is distributed
by the National Optical Astronomy Observatories, which are operated by
the Association of Universities for Research in Astronomy, Inc., under
cooperative agreement with the National Science Foundation.}. For
wavelength calibration, we used a Th-Ar spectrum.  After removing
telluric lines by dividing the object spectra by the standard star
spectrum (see Sect. 2.1), we directly fitted the continuum with a
third-order cubic spline function.  To increase signal-to-noise (S/N),
every 0.1~\AA-wide segments are smoothed and resampled.

\subsection{Removal of Telluric Lines}
Once we extract one-dimensional spectra of three target stars, we need
to remove telluric lines by using a standard star spectrum.  Because
DIBs, whose profiles are usually very broad and shallow, are easily
buried in strong telluric lines, a careful removal is necessary.

First, we normalize the standard star spectrum, leaving only telluric
features because rotationally broadened stellar lines are fitted out
as continuum.  Because the object stars are observed under slightly
different airmass from that of the standard star, we have to correct
telluric lines according to their optical depth, as
\begin{equation}
  f_{sky}^{corr} = exp\left(\ln(f_{sky})\times\frac{secz({\rm obj})}{secz({\rm ss})}\right),
\end{equation}

where $f_{sky}$ and $f_{sky}^{corr}$ are raw and corrected normalized
spectra of telluric lines, respectively. The sec$z$(obj,ss) is the
airmass toward the object or standard star.  Here, we assume a plane
parallel atmosphere, because sec$z$ toward our targets (including a
standard star) are all close to 1 (i.e., close to zenith) and because
object/standard stars were all observed close in time to each other.
Finally, we obtain telluric free spectra by dividing the spectra of
object stars by the airmass-corrected telluric line spectrum.

We observed only one standard star for the telluric line removal, and
its spectral type is slightly different from those of the object
stars.  Nonetheless, the photospheric features from the standard star
do not affect the final object spectra significantly, because their
line profiles are all broad and shallow and they should be fitted out
while continuum fitting.  On the other hand, any photospheric features
from the object stars will still remain after dividing by the standard
star spectrum.  We investigate these in detail in Section 4.3.

An observation log is given in Table~1, in which we list target name,
coordinate (RA and Dec), apparent visual magnitude, spectral type,
total exposure time, signal-to-noise (S/N) ratio per pixel after
sampling, averaged sec$z$, degree of reddening, dust extinction in the
visual band, effective temperature, and surface gravity.  Normalized
spectra of our three target stars are presented in Figure~2. We show
only five DIB regions, which we discuss below.

\section{Identifications of DIBs}
At first we search for the two DIBs at $\lambda$ $\sim$ 9577~\AA\ and
9632~\AA\ (i.e., candidates for $C_{60}^+$ absorption) in the spectra
of three target stars.  Among these, detections have been reported
only toward HD~37022 (e.g., \citealt[]{jen94,foi97,mou99}).
We use all absorption features that are detected with $\geq$ 5$\sigma$
level.  Both DIBs are detected toward the two stars within the nebula,
while no features are detected toward HD~37150, with an equivalent
width detection limit of EW = 7.0~m\AA\ and 5.5~m\AA\ for DIB~9577 and
9632, respectively.  Based on laboratory measurements, \citet{ful93}
present wavelengths of 13 vibrational-mode absorption lines of
$C_{60}^+$ (in neon matrix) at 8323~\AA\ -- 9642~\AA\ (in air), of
which 7 are covered by our Subaru spectra.  Our identification results
for the three target spectra are summarized in Table~2. Following
column (1) with ID name, columns (2) and (3) give vacuum wavenumbers
and air wavelengths of expected $C_{60}^+$ features, columns (4), (5),
(6), and (7) are flux-weighted central wavelengths, EW, wavelength
shift from the laboratory measurement, and full-width at half maximum
(FWHM) of DIBs toward HD~37022. Columns (8), (9), (10) and (11) are
identical to (4) -- (7) but toward HD~37041, column (12) shows the
5$\sigma$ detection limit of DIBs toward HD~37150.
We did not correct for contamination from the stellar \ion{Mg}{2}
lines when we measure equivalent widths of DIB~9632, because their
equivalent widths are no more than 50~m\AA\ for spectral types of O7
-- B3I (\citealt{gal00}; see further discussion in Section 4.3). All
of our targets and the standard star lie in this category.
Interestingly, three DIBs are newly identified as possible $C_{60}^+$
lines, and detected only toward stars in the nebula.
The equivalent width and the FWHM of each DIB are measured and
presented in Table~2, although the latter has much uncertainty.
Because of the presence of substructure (see Section 4.2) in the DIB
and the contamination with strong telluric lines, it is not
straightforward to measure widths.  In Figure~2, we shaded the
wavelength regions around the five DIBs, over which we carried out the
measurements.

All target stars are affected by both the interstellar and the local
extinction.  The degree of reddening, $E(B-V)$, listed in Table~1 is
the difference between the observed $B-V$ color and the expected $B-V$
color, where the expected color is estimated from the predictions of
\citet{bes98} by using the effective temperature and surface gravity
measurements from \citet{bro94}.  We also list extinction values from
\citet{bro94}.  The reddening is significant toward HD~37022 in
Trapezium, and also toward HD~37041, which has only a small offset of
$\sim$2.4$^{\prime}$ ($\sim$0.3~pc) from HD~37022.  Both stars lie in
highly ionized central region of the Orion Nebula.  On the other hand,
HD~37150, whose distance from Trapezium is $\sim$20$^{\prime}$
($\sim$2.5~pc), has small reddening.  Thus, our target selection
enables us to trace two regions of the Orion Nebula with very
different physical conditions.

\section{Discussion}
We can draw several inferences on the DIB carriers in the Orion Nebula
from our observations; physical condition with regard to their
internal structure, and correlation with dust abundance.  We have also
found additional candidates of $C_{60}^+$ DIBs. We discuss all these
below.

\subsection{Physical Conditions of the DIB Carriers}
Our targets are distributed at the center or the edge of the Orion
Nebula, which enables us to compare the detectability of $C_{60}^+$
DIBs in two extreme physical conditions. The regions toward HD~37022
and HD~37041 are strongly UV-irradiated by the hot Trapezium stars,
while the ionization condition toward HD~37150 is probably much lower.
Visual extinction is also very different --- the former two are
severely, and the latter is only slightly, reddened.  DIBs are known
to have a strong correlation with neutral hydrogen column density
(i.e., gas abundance; \citealt[]{wel06}) as well as dust extinction
(i.e., dust abundance; \citealt[]{her93}).  Local UV radiation may
enhance the 9577/9632 DIBs \citep{foi97}.  We detect these DIBs only
toward two stars at the strongly UV-irradiated regions with much
amount of dust, which is consistent with past results.  This trend can
be explained by a ``skin effect'' --- the effective shielding of UV
radiation by the inner layers of dust in the nebula, as proposed in
\citet{cam97}.
\citet{cam97} also measured the complete DIB spectrum of HD~37022, and
discovered that almost all optical DIBs (except for DIBs at $\lambda$
= 5780~\AA\ and 6284~\AA, as shown in \citealt{jen94}) were very weak
or not detected toward this extremely high UV-irradiated region.  It
is likely that the carriers of these DIBs are dissociated under such
high ionizing radiation.  On the other hand, 5780/6284 DIBs as well as
9577/9632 DIBs show a clear correlation of their absorption strengths
with their local ionization levels, which means these carriers prefer
an ionized phase (i.e., $C_{60}^+$ with ionization potential (IP) =
11.3~eV for the 9577/9632 DIBs) to a neutral phase (i.e., $C_{60}$
with IP = 7.6~eV) \citep{foi97}.
Thus, we confirm the 9577/9632~\AA\ DIB carriers have small-scale
(i.e., on a few parsec scale) spatial fluctuation in a single cloud
(i.e., the Orion Nebula), while \citet{gal00}, already noted a
Galactic scale fluctuation.  We reconfirm that the detection of the
9577/9632 DIB require large amount of dust as well as strong UV
radiation.

\subsection{Internal Substructure in the DIBs}
The 9577/9632 DIBs were previously treated as a single Gaussian
profile (e.g., \citealt[]{foi97}) and their equivalent widths are also
measured using this assumption (e.g., \citealt[]{gal00}).  However,
the DIBs toward the Orion Nebula clearly show asymmetrical profiles
with possible multiple components in our high resolution spectra.
Actually, this asymmetrical profile can be seen in the past papers,
especially for DIB~9632 (e.g., \citealt[]{foi97}), although it was not
discussed in detail.  \citet{ehr96} resolved three optical DIBs at
5797, 6379, and 6613~\AA\ into two or three subcomponents, following
the first discovery of substructure by \citet{wes88a,wes88b}.  They
conclude that rotational contours probably play a major role in this
internal structure, following \citet{dan76}, although there could be
small contributions from other sources (e.g., instrumental line
spread, thermal broadening, gas turbulence, and intramolecular
vibration rotation energy transfer).

Here, we would like to suggest an additional contributor to the
broadening that we observe; i.e., the existence of discrete absorbers
along our sight lines.  \citet{ode93} took high resolution spectra of
four Trapezium stars (including HD~37022 and HD~37041), and discovered
internal substructures with multiple absorption components in
\ion{Ca}{2} and \ion{Na}{1} absorption profiles. The typical velocity
distributions of these lines are $\sim$30--50~\kms\ and up to 60~\kms,
which is one third of the FWHM ($\sim$130~\kms) of the 9577/9632 DIBs
we observed. If each velocity component seen in \ion{Ca}{2} and
\ion{Na}{1} gives rise to the DIB profiles, these would result in a
substantial contribution to the DIB broadening.
We performed a simple test for this scenario as follows. \citet{foi97}
measured the FWHM of the 9577/9632~DIBs toward three stars (including
one of our targets, HD~37022), and obtained a slightly larger line
width toward HD~37022 (FWHM $\sim$4~\AA) as compared to the line
widths toward two other stars (FWHM $\sim$2.9~\AA).  This difference
can be reproduced by adding an internal velocity dispersion of
$\sim$80~\kms\ to a DIB cloud toward HD~37022, which is possible
because multiple \ion{Ca}{2} and \ion{Na}{1} clouds are distributed
within a similar velocity width, $\leq$60~\kms, as described
above. Our results suggest that several discrete DIB clouds contribute
non-negligibly to the total DIB line width.
However, it should be noted that the major source of DIB broadening is
likely to be intrinsic to the molecular structure of the carriers,
because (i) such an internal structure is seen even if corresponding
atomic lines have single Gaussian profiles at the same velocity (e.g.,
Figure~1 of \citealt[]{gal08}), and because (ii) DIBs have similar
characteristic profiles toward various sightlines (e.g., Figure~2 of
\citealt[]{gal08}). Both of these support the idea that the main
source of DIB broadening is molecular in origin.

For further investigation of the internal structure of the 9577/9632
DIBs, high quality spectra with higher S/N ratio would be required.

\subsection{Other Possible $C_{60}^+$ DIBs}
Although \citet{foi94} detected two DIBs within 10~\AA\ of the
laboratory-measured positions of two $C_{60}^+$ absorption features
\citep{ful93}, this identification was questioned by \citet{jen97}
because two other $C_{60}^+$ absorption lines at 9366~\AA\ and
9419~\AA\ are not detected simultaneously.  Thus, this identification
is not yet a matter of consensus. Additional evidence is required for
further discussion.
Using our Subaru spectra with relatively wide wavelength coverage from
8700~\AA\ to 1~$\mu m$, we attempt to confirm the existence of 7 of 9
$C_{60}^+$ lines (in addition to the 9577/9632 DIBs) from
\citet{ful93}.  Unfortunately, the 9372/9429~\AA\ lines in question
(at $\lambda$ $\sim$ 9366 and 9419~\AA, according to
\citealt[]{jen97}) are on an echelle order gap or wavelength region
that is hopelessly affected by telluric lines.  Among the other 5
lines, three at $\lambda$ $\sim$ 9258, 9210, and 9017~\AA\ are clearly
detected with $\leq$ 10~\AA\ difference from the laboratory
measurements, but one at $\lambda$ $\sim$ 9154~\AA\ is not seen with
$\geq$ 5$\sigma$ level toward HD~37022 and HD~37041 (see Table~2). We
cannot search for a line at $\lambda$ $\sim$ 8954~\AA\ because it is
on an echelle order gap. No lines are detected toward HD~37150.

In order to rule out the possibility that these five DIBs arise in the
stellar photosphere, we synthesized stellar spectra using model
atmospheres with the ATLAS9 and SYNTHE codes \citep{kur05,sbo05}.  We
used model parameters such as stellar rotational velocity, effective
temperature, and surface gravity that are appropriate to each star (as
listed in Table~1) from the literature \citep{bro94}.  The model
spectra are overlaid on the observed spectra and the most prominent
absorption features are named in Figure~2.  Below, we discuss each
DIB.

\begin{description}

\item[DIB~9633. ---] This is one of the two DIBs that have been
proposed to be associated with $C_{60}^+$.  Although it is partially
blending with \ion{Mg}{2} line from stellar photosphere, the
\ion{Mg}{2} line center is clearly shifted blueward from the center of
the DIB.  Moreover, the \ion{Mg}{2} line is weak compared to the
DIB. This is the case especially for type~O stars.  This was one of
the first DIBs discovered in the near-infrared.

\item[DIB~9577. ---] Another DIB whose origin was suggested to be
$C_{60}^+$.  There are no narrow stellar intrinsic lines within
10~\AA\ of the DIB, although the DIB is positioned at the red wing of
a very broad \ion{H}{1} line at $\lambda$ = 9546.0~\AA.  The detection
of this DIB is highly secure.

\item[DIB~9258. ---] We have discovered this DIB, and discuss it here
for the first time. Although it is located at the edge of an echelle
order, the detection itself is secure because line features toward two
different stars (i.e., HD~37022 and HD~37041) are almost identical.
Their origin is neither stellar photospheric, nor due to any
instrumental data defect.  The feature is only seen toward the
Trapezium stars, as noted for the two DIBs mentioned above.  An
absorption feature at $\Delta \lambda$ = $\sim$5 -- 10~\AA\ toward
HD~37150 is probably caused by the \ion{N}{1} line complex within the
stellar photosphere, which is more prominent for type~B stars than
type~O stars (see Figure~2).  A mild inclination of the model continua
is due to a strong \ion{Mg}{2} line at $\lambda$ = 9244.4~\AA.

\item[DIB~9210. ---] Only this DIB was significantly blending with
stellar intrinsic features of \ion{He}{1} at $\lambda$ = 9210.3~\AA\
and 9213.2~\AA\ and \ion{Mg}{2} at $\lambda$ = 9218.3~\AA.  However,
the flux-weighted centers are shifted slightly redward from the line
center of this DIB.  The whole range of the observed feature cannot be
reproduced only by the stellar photospheric lines, and much of the
absorption must originate in a DIB. The equivalent width and line
width of this DIB listed in Table~2 are only upper limits.  The
inclined continuum of the model spectra is due to a strong \ion{H}{1}
line at $\lambda$ = 9229.0~\AA.

\item[DIB~9017. ---] This is the weakest one among the five DIBs we
detected in our spectra.  The DIB is detected only toward the
Trapezium stars like the other DIBs, which also supports this
discovery.  A strong \ion{H}{1} line is located at $\lambda$ =
9014.9~\AA\ near the DIB, but it should not affect the DIB because of
its broad and smooth line profile (i.e., it will be fitted out during
spectrum normalization).

\end{description}

Interestingly, all detected DIBs are blue-shifted from the laboratory
measurements by 4--8~\AA\ ($\Delta v$ $\sim$ 150--200~\kms); this is
the same trend as seen in the 9577/9632 DIBs. This blue-shift is not
due to any outflow of DIB carriers from the Orion Nebula, because the
corresponding \ion{Ca}{2} and \ion{Na}{1} absorption lines are
blue-shifted only up to $|\Delta v|$ $<$ 20~\kms\ in heliocentric
velocity, about one order of magnitude smaller than the DIB velocity
shift \citep{ode93}.
We suspect this is a systematic trend due to differences in physical
conditions between the gas-phase inter stellar medium and the
laboratory neon matrix measurement, as found for other molecules like
naphthalene (C$_{10}$H$_8$$^+$; \citealt[]{sal92}), coronene and
ovalene (C$_{24}$H$_{12}$ and C$_{32}$H$_{14}$;
\citealt[]{ehr92}). This wavelength shift is also estimated to be
smaller in the near-infrared band at $\lambda$ $\sim$ 1~$\mu$m
($\Delta k$ $\sim$ 5 -- 10~\cm, which is almost same value as our
results in Table~2) than that in the optical band ($\sim$ 20 --
50~\cm; \citealt{gal00}).
The measurement in matrix would also change the relative strengths of
absorption lines \citep{ful93}.  This may explain our result that the
$C_{60}^+$ lines at $\lambda$ $\sim$ 9528, 9210, and 9012~\AA, whose
strengths are very weak (by a factor of $\sim$ 10 compared to the
stronger 9577/9632 lines) in the laboratory measurements
\citep{ful93}, are detected with almost same equivalent widths as
those of the 9577/9633 DIBs in our spectra.
If these DIBs are really arising from the same molecule, their line
widths also would be expected to be similar \citep{mai94}. The FWHMs
of the three newly detected DIBs at $\lambda$ $\sim$ 9258, 9210, and
9017~\AA\ are $\geq$3~\AA, $\sim$6.5~\AA, and $\sim$2.5~\AA,
respectively, which are scattered around those of the 9577/9632 DIBs,
$\sim$4.5~\AA.  This could be due to contamination of other atomic
lines (for DIB~9210) or a failure in detecting weak profile wings (for
DIB~9017).  More precise measurements for $C_{60}^+$ in laboratory
without using neon/argon matrix are desirable.

\subsection{Correlation of the DIB Strength with Dust Abundance}
From their DIB survey, targeting the Milky Way, the LMC, and the SMC,
\citet{wel06} have found a clear correlation between the strength of
DIBs and the \ion{H}{1} column density along the line of sight.
Following this result, \citet{law08} embarked on a survey of the 5780
DIB in Damped Ly$\alpha$ (DLA) systems (with $\log$(\ion{H}{1}) of
$\geq$ 20.3 [\cmm]) at cosmological distances, and discovered one
system with two DIBs at $\lambda_{rest}$ of 5705~\AA\ and 5780~\AA\
with moderate reddening DLA system (E(B$-$V) = 0.23) at $z$ = 0.5
\citep{yor06}. So far, this is the highest redshift at which DIBs have
been detected.  This detection would imply that DIB carriers are
distributed universally, and their abundance probably has a strong
correlation with dust as well as gas abundance.
Similar correlations have also been noted for the 9577/9632 DIBs
\citep{gal00}, and our results are consistent with these (see
Figure~3).  Although our measured 9577~DIB strengths lie slightly
above from the expected value, this may be because the ratio of
total-to-selective extinction, $R_V$ = $A_V / E(B-V)$, is larger
toward stars in the Orion Nebula, as shown in Table~1.

The equivalent width ratio of these DIBs should remain unchanged by
the absorber's physical conditions if they all have the same carrier
\citep{mai94}.  \citet{ful93} estimated the relative strength of the
9577/9632 DIBs to be $\sim$ 1.5 in a neon matrix, but with an
uncertainty over the range of 1.5 to 2.0, depending on production
conditions.  On the other hand, this ratio is close to 1 in our
observation toward two stars in the Orion Nebula (see Figure~3). This
is instead consistent with other measurements.  \citet{jen97} and
\citet{gal00} also found the ratio of close to 1. If it can be
confirmed that the gas phase measurement gives a smaller intensity
ratio of close to unity in the laboratory, the identification of these
DIBs with $C_{60}^+$ would be strengthened.

\section{Summary}
We have investigated the detectability of the 9577/9632 DIBs that are
candidates of $C_{60}^+$ absorption features. We obtained spectra of
two stars in the Orion Nebula and one star at the edge of the nebula.
These DIBs are detected only toward the two stars in the nebula.

Furthermore, for the first time, we have detected three other DIBs
that could also arise in $C_{60}^+$.  The wavelengths of these
features matches the laboratory values to within 10~\AA. With our
high-resolution spectra ($R$ = 120,000), we found internal structures
in the DIBs with two or three components. If our sight-lines really
trace multiple clouds, those could partially contribute to the line
broadening of the DIBs, though the main cause is likely to be relate
to molecular transitions.  The DIB strength and dust reddening from
our data are consistent with correlations seen by previous studies,
which means that we are probably observing the same DIB carrier. For
further investigation, we need (i) spectra with higher S/N ratio, (ii)
precise measurements of $C_{60}^+$ in the laboratory without using
matrix, and (iii) spectra taken from the space (e.g., using Hubble
Space Telescope) to detect the DIBs without any contamination from the
telluric lines.

\acknowledgments TM and AH acknowledge support from the Special
  Postdoctoral Research Program and DRI Research Grant of RIKEN.  PG
  acknowledges support from the Special Foreign Postdoctoral Research
  Program of RIKEN.  TM acknowledges support from the Sumitomo
  Foundation (070380).  We would like to thank the anonymous referee
  for their valuable comments.

\clearpage


\begin{deluxetable}{cccccccccccc}
\tabletypesize{\scriptsize}
\setlength{\tabcolsep}{0.05in}
\setcounter{table}{0}
\tablecaption{Log of Observations \label{t1}}
\tablewidth{0pt}
\tablehead{
\colhead{Star}          &
\colhead{RA$^a$}        & 
\colhead{Dec$^a$}       & 
\colhead{$m_V$}         & 
\colhead{Spectral Type} & 
\colhead{Exposure}      &
\colhead{S/N$^b$}       &
\colhead{$\langle$sec$z$$\rangle$$^c$} &
\colhead{E(B$-$V)$^d$}  & 
\colhead{$A_V$$^e$}     &
\colhead{$\log$($T_{eff}$)$^f$}   & 
\colhead{$\log$(g)$^g$} \\
\colhead{}              &
\colhead{(h:m:s)}       & 
\colhead{(d:m:s)}       &
\colhead{(mag.)}        & 
\colhead{}              &
\colhead{(sec)}         & 
\colhead{(pix$^{-1}$)}  &
\colhead{}              &
\colhead{(mag.)}        & 
\colhead{(mag.)}        &
\colhead{(K)}           & 
\colhead{(cm~s$^{-2}$)} \\
\colhead{(1)}           &
\colhead{(2)}           &
\colhead{(3)}           &
\colhead{(4)}           &
\colhead{(5)}           &
\colhead{(6)}           &
\colhead{(7)}           &
\colhead{(8)}           &
\colhead{(9)}           &
\colhead{(10)}          &
\colhead{(11)}          &
\colhead{(12)}          
}
\startdata
HD~37022  & 05 35 16.5 & -05 23 23 & 5.13 & O6pe    & 240 & 200 & 1.129 & 0.33 & 1.74 & 4.64 & 4.80 \\
HD~37041  & 05 35 22.9 & -05 24 58 & 5.08 & O9.5Vpe & 240 & 211 & 1.137 & 0.21 & 1.12 & 4.50 & 4.18 \\
HD~37150  & 05 36 15.0 & -05 38 53 & 6.51 & B3Vv    & 840 & 139 & 1.185 & 0.03 & 0.05 & 4.31 & 4.30 \\
HD~37397  & 05 38 13.7 & -01 10 09 & 6.81 & B2V     & 840 &     & 1.113 & 0.05 & 0.09 & 4.28 & 4.47 \\
\enddata
\tablenotetext{a}{J~2000.0 coordinates.}
\tablenotetext{b}{Signal to noise ratio per 0.1~\AA\ around wavelength
   of our interest, i.e., $\lambda$ $\sim$ 9600~\AA, after removing
   telluric lines.}
\tablenotetext{c}{Average airmass to the target during exposure,
   assuming plane-parallel atmosphere.}
\tablenotetext{d}{Degree of reddening, a difference between the
   observed $B-V$ color and the expected $B-V$ color, where the
   expected color is estimated from the predictions of \citet{bes98}.}
\tablenotetext{e}{Dust extinction in visual band from the literature
  \citep{sim06,bro94}.}
\tablenotetext{f}{Logarithm of effective temperature from
   \citet{bro94}.}
\tablenotetext{g}{Logarithm of surface gravity from \citet{bro94}.}
\end{deluxetable}

\begin{deluxetable}{ccccccccccccccc}
\tabletypesize{\scriptsize}
\tablecaption{Possible identification of DIBs with $C_{60}^+$ \label{t2}}
\tablewidth{0pt}
\tablehead{
\colhead{}                       &
\multicolumn{2}{c}{$C_{60}^+$}   &
\colhead{}                       &
\multicolumn{4}{c}{HD~37022}     &
\colhead{}                       &
\multicolumn{4}{c}{HD~37041}     &
\colhead{}                       &
\colhead{HD~37150}               \\
\noalign{\vskip 3pt}
\cline{2-3}
\cline{5-8}
\cline{10-13}
\cline{15-15}
\noalign{\vskip 3pt}
\colhead{ID}                         &
\colhead{$k_{vac}$$^a$}              &
\colhead{$\lambda_{air}$$^b$}        &
\colhead{}                           &
\colhead{$\lambda_c$}                &
\colhead{EW$^c$}                     &
\colhead{$\Delta\lambda$$^d$}        &
\colhead{FWHM$^e$}                   &
\colhead{}                           &
\colhead{$\lambda_c$}                &
\colhead{EW$^c$}                     &
\colhead{$\Delta\lambda$$^d$}        &
\colhead{FWHM$^e$}                   &
\colhead{}                           &
\colhead{EW$^c$}                     \\
\colhead{}                       &
\colhead{(\cm)}                  &
\colhead{(\AA)}                  &
\colhead{}                       &
\colhead{(\AA)}                  & 
\colhead{(m\AA)}                 & 
\colhead{(\AA)}                  &
\colhead{(\AA)}                  &
\colhead{}                       &
\colhead{(\AA)}                  & 
\colhead{(m\AA)}                 &
\colhead{(\AA)}                  & 
\colhead{(\AA)}                  &
\colhead{}                       &
\colhead{(m\AA)}                 \\
\colhead{(1)}                    &
\colhead{(2)}                    &
\colhead{(3)}                    &
\colhead{}                       &
\colhead{(4)}                    &
\colhead{(5)}                    &
\colhead{(6)}                    &
\colhead{(7)}                    &
\colhead{}                       &
\colhead{(8)}                    &
\colhead{(9)}                    &
\colhead{(10)}                   &
\colhead{(11)}                   &
\colhead{}                       &
\colhead{(12)}                   
}
\startdata
A     & 10368 & 9642.4 & & 9633.2 &  95.7$\pm$2.8 & $-$9.2 &    4.2 & & 9633.0 &  98.2$\pm$3.1     & $-$9.4 &    4.5 & & $<$5.5 \\
B     & 10435 & 9580.5 & & 9577.0 & 103.5$\pm$3.0 & $-$3.5 &    4.4 & & 9577.2 & 131.8$\pm$3.5$^f$ & $-$3.3 &    5.4 & & $<$7.0 \\
C$^g$ & 10603 & 9428.7 & & ...    &  ...          & ...    &    ... & & ...    &  ...              & ...    &    ... & & ...    \\
D$^h$ & 10667 & 9372.1 & & ...    &  ...          & ...    &    ... & & ...    &  ...              & ...    &    ... & & ...    \\
E$^i$ & 10792 & 9263.6 & & 9258.3 &  88.6$\pm$2.4 & $-$4.7 & $>$3.4 & & 9258.3 &  96.4$\pm$2.5     & $-$5.3 & $>$3.3 & & $<$6.0 \\
F     & 10845 & 9218.3 & & 9209.7 & 129.2$\pm$3.4 & $-$8.6 &    6.6 & & 9209.7 & 121.8$\pm$3.5     & $-$8.6 &    6.5 & & $<$6.5 \\
G     & 10922 & 9153.5 & & ...    &        $<$5.0 & ...    &    ... & & ...    &        $<$4.0     & ...    &    ... & & $<$7.0 \\
H     & 11082 & 9021.1 & & 9017.0 &  31.0$\pm$1.7 & $-$4.1 &    2.3 & & 9017.8 &   9.7$\pm$2.7$^j$ & $-$3.3 &    2.8 & & $<$5.5 \\
I$^h$ & 11165 & 8954.1 & & ...    &  ...          & ...    &    ... & & ...    &  ...              & ...    &    ... & & ...    \\
\enddata
\tablenotetext{a}{Vacuum absorption wavenumber of $C_{60}^+$ in neon
   matrix from \citet{ful93}.}
\tablenotetext{b}{Air absorption wavelength of $C_{60}^+$, converted
  from \citet{ful93}.}
\tablenotetext{c}{Equivalent width with 1$\sigma$ error of DIBs, or
   5$\sigma$ detection limit if not detected.}
\tablenotetext{d}{Difference in wavelength of DIB from the laboratory
   measurement.}
\tablenotetext{e}{Full-Width-Half-Maximum of DIB in angstrom.}
\tablenotetext{f}{Equivalent width is measured at up to $\lambda$ =
  9579.3\AA, because contamination of telluric lines is severe at the
  longer wavelength region.}
\tablenotetext{g}{This DIB is significantly affected by strong
  telluric lines.}
\tablenotetext{h}{This DIB is not covered in our spectra because it is
   located at an echelle order gap.}
\tablenotetext{i}{This DIB is located at the edge of an echelle order
   (see Figure~2), so we can place only upper limit of $\lambda_c$ and
   lower limit of EW(DIB).}
\tablenotetext{j}{This DIB is detected with 3.6$\sigma$ level.}
\end{deluxetable}
\clearpage


\begin{figure}
 \begin{center}
  \includegraphics[width=12cm,angle=0]{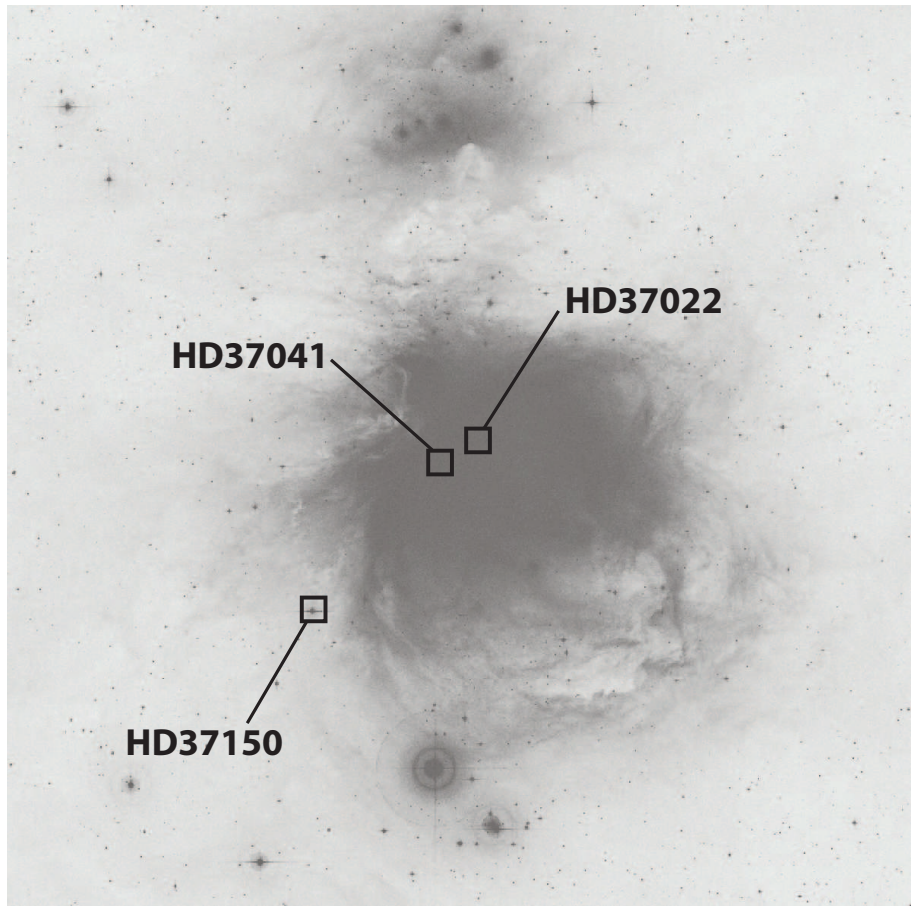}
 \end{center}
 \caption{R-band image of the Orion Nebula with the field of
 90$^{\prime}$ $\times$ 90$^{\prime}$, retrieved from the Digital Sky
 Survey.  The locations of our target stars (HD~37022, HD~37041, and
 HD~37150) are marked with open squares.}
\end{figure}

\begin{figure}
 \begin{center}
  \includegraphics[width=12cm,angle=270]{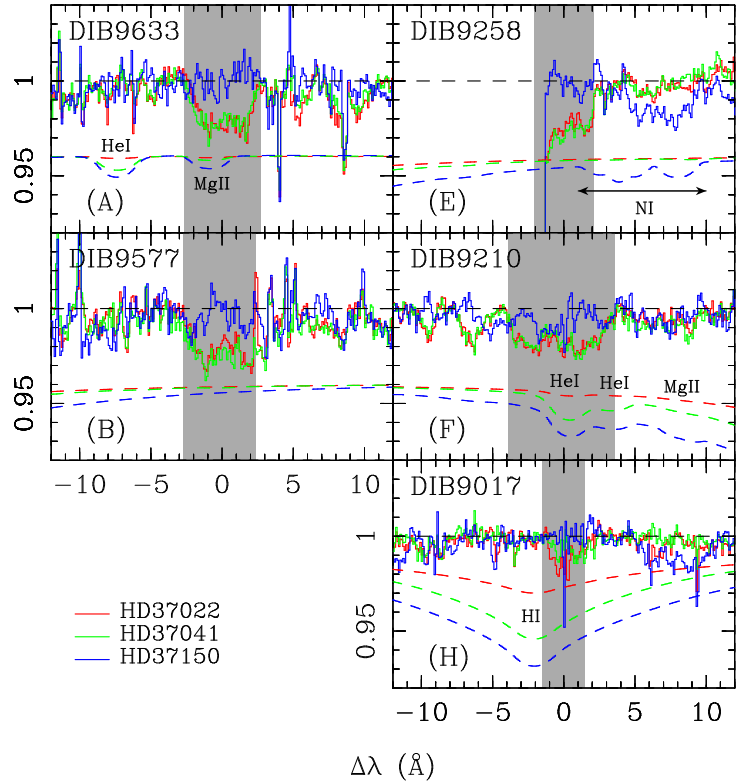}
 \end{center}
 \caption{Spectra of three target stars around the five detected DIBs
 at $\lambda$ = 9633, 9577, 9258, 9210, and 9017~\AA\ (solid lines).
 These DIBs are detected toward HD~37022 and HD~37041 with $\geq$
 5$\sigma$ level, except for the DIB~9017 toward HD~37041 whose
 detection level in $\sim$ 3.6$\sigma$. DIB~9259 is located at the
 edge of an echelle order. No DIBs are detected toward HD~37150. We
 also overlay synthesized stellar spectra (dashed lines) using model
 atmosphere with ATLAS9 and SYNTHE \citep{kur05,sbo05}.  Model spectra
 are all compressed vertically by a factor of 4 and their continuum levels
 are shifted down to 0.96, except for DIB~9017, where the continuum
 level is 0.99.  Strong absorption features in the model spectra are
 labeled.}
\end{figure}

\clearpage
\begin{figure}
 \begin{center}
  \includegraphics[width=15cm,angle=270]{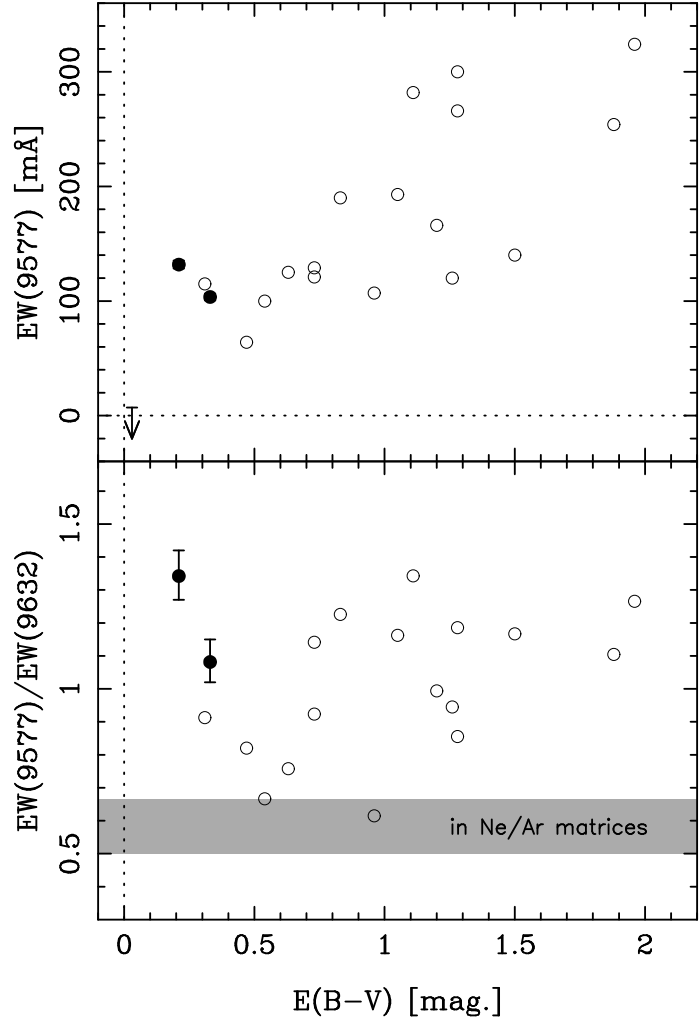}
 \end{center}
 \caption{Correlation between the equivalent width of DIB~9577 and
 $E(B-V)$ (top) and between the equivalent width ratio of DIB~9577 to
 DIB~9632 and $E(B-V)$ (bottom). Open circles are Galactic DIBs from
 \citet{gal00}, and filled circles and a downward arrow are our
 results toward the Orion Nebula. The laboratory measurement using Ne
 or Ar matrix is indicated by the shaded region \citep{ful93}.  Our
 results are consistent with the correlation already seen toward other
 Galactic stars, but not with the laboratory measurement.}
\end{figure}

\end{document}